\documentclass[twocolumn]{jpsj2} 
\title{Superexchange Interaction and Magnetic Moment in \\ Antiferromagnetic High-$T_c$ Copper-oxide Superconductors}

\author{Yoshio \textsc{Kitaoka}\thanks{E-mail: kitaoka@mp.es.osaka-u.ac.jp}, Hidekazu \textsc{Mukuda}, Sunao \textsc{Shimizu}}
\inst{Department of Materials Engineering Science, Osaka University, Toyonaka, Osaka 560-8531, Japan}
\recdate{\today}

\abst{
Extensive $^{63}$Cu-NMR studies on multilayered high-$T_c$ cuprates have deduced the following results;
(1) Antiferromagnetic (AFM) moment $M_{\rm AFM}$ is decreased with doping, regardless of the number of CuO$_2$ layers $n$, and collapses around a carrier density $N_h\sim$ 0.17. (2) The AFM ordering temperature is enhanced as the out-of-plane coupling $J_{\rm out}$ increases with increasing $n$. (3) The in-plane superexchange $J_{\rm in}$ is invariant with doping, but is even increased. (4) The dome shape of $T_c$ from the underdoped to the overdoped regime with a maximum $T_c$ at $N_h\sim$ 0.22 does not depend on $n$, but its maximum value of $T_c$ seems to depend on $n$ moderately. The present results  strongly suggest that the AFM interaction plays the vital role as the glue for the Cooper pairs, which will lead us to a genuine understanding of why the $T_c$ of cuprate superconductors is so high.} 

\kword{Phase Diagram, High-$T_c$ superconductivity, Antiferromagnetism, Superexchange}

\begin{document}
\maketitle
Despite more than 22 years of intensive research, an origin of high-temperature copper-oxides superconductivity (HTSC) has not been well understood yet. The HTSC emerges on a CuO$_2$ plane when an antiferromagnetic
Mott insulator is doped with mobile carriers. A strong relationship between antiferromagnetism (AFM) and superconductivity (SC) is believed to be a key to understand the origin of the remarkably high-SC transition.\cite{Chen,Giamarchi,Inaba,Anderson1,Lee1,Zhang,Himeda,Kotliar,Paramekanti1,Lee2,Demler,Paramekanti2,Shih1,Shih2,Senechal,Ogata,Kyung,Pathak}
Site-selective $^{63}$Cu-NMR studies on multilayered cuprates revealed that the square-type inner CuO$_2$ planes (IPs) exhibit homogeneous hole doping, since the IPs are farther from the charge reservoir layers and the disorder introduced along with the chemical substitution in it is effectively shielded on a pyramid-type outer CuO$_2$ plane (OP). As a result, ideally flat CuO$_2$ planes are realized, especially at IPs.\cite{Tokunaga,Kotegawa2001,Kotegawa2004,Mukuda2006,Mukuda2008,Shimizu,Shimizu1,Shimizu2}  Thus, multilayered copper oxides provide us with the opportunity to research the characteristics of the disorder-free CuO$_2$ planes which are coupled each other.

On the basis of two-layered ($n$ = 2), four-layered ($n$ = 4), and five-layered ($n$ = 5) copper oxides, respective figures 1(a), (b) and (c) reveal the phase diagrams of AFM and SC, where $T_c$ and $T_N$ are plotted against carrier density $N_h$, for the $n=2$, $n=4$, and $n=5$ compounds.\cite{Mukuda2006,Mukuda2008,Shimizu,Shimizu1,Shimizu2,Shimizu3} Note that the phase diagram of the $n=2$ compounds does not reveal an AFM order around $N_h < 0.15$,\cite{Shimizu1,Shimizu3} resembling the well-established phase diagram of YBCO.\cite{YBCO} However, the AFM phase in the $n=4$ compounds, which uniformly coexists with SC, exists up to $N_h\sim$ 0.15 being an AFM quantum critical point (QCP) of the $n=4$ compounds.\cite{Shimizu2} When $n$ increases from $n$ = 4 to 5, the QCP moves to higher hole doped region, $N_h\sim$ 0.17.\cite{Mukuda2008} This result suggests that an interlayer magnetic coupling $J_{\rm out}$ becomes stronger with increasing $n$, which stabilizes the AFM order. The phase diagram of AFM and SC in multilayered systems, especially in the  $n=4$ and 5 compounds, are remarkably different from the well-established ones for LSCO ($n=1$) and YBCO ($n=2$), where the AFM order totally collapses by doping very small amount of holes with $N_h\sim$ 0.02 \cite{LSCO} and $N_h\sim$ 0.055,\cite{YBCO} respectively. The QCPs for $n=4$ and 5 compounds are located at the doping levels higher than those for $n=1$ and 2 compounds, thereby the AFM uniformly coexists with SC. These results ensure that decreasing $n$ makes $J_{\rm out}$ weaker and an AFM ordering temperature depends on $J_{\rm out}$ significantly.
\begin{figure}[h]
\centering
\includegraphics[width=6.5cm]{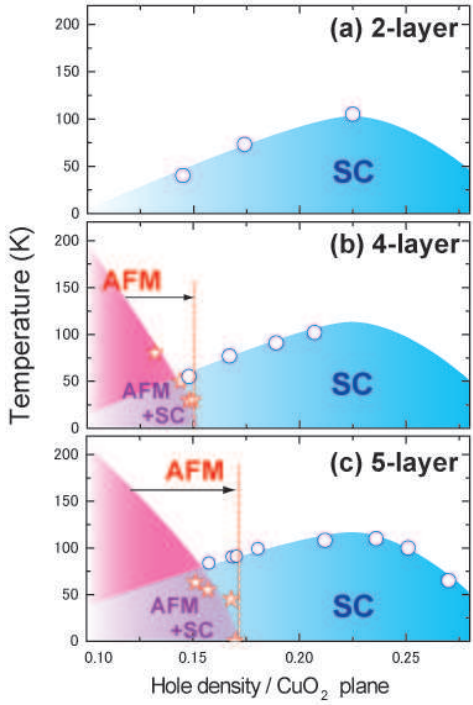}
\caption[]{\footnotesize (Color online) The phase diagrams of AFM and SC; (a) the $n=2$ compounds \cite{Shimizu1,Shimizu3}, (b) the $n=4$ compounds \cite{Shimizu2}, and (c) the $n=5$ compounds \cite{Mukuda2008}. Note that the QCP for the $n=2$ compounds is lower than $N_h\sim$ 0.15 at least and $N_h$ = 0.15 and 0.17 for the $n=4$ and 5 compounds, respectively.}
\end{figure}
\\
Consequently, we have concluded that the uniform mixing of AFM and SC is a general property inherent to a single CuO$_2$ plane in an underdoped regime of HTSC.\cite{Mukuda2006,Mukuda2008,Shimizu,Shimizu1,Shimizu2} This conclusion is corroborated by the ARPES result \cite{Chen2} on the $n$=4 compound; it was found that the two Fermi sheets of the IP and OP are observed, and that the SC gap opens at the IP and OP below $T_c$= 55 K where the AFM order takes place below $T_N=80$ K. Note that $T_c$ is almost independent of $n$ when $n \ge 4$ . This is because the OP is responsible for the onset of bulk SC with its maximum $T_c\sim 100$ K in these compounds. The highest $T_c$ was observed around 133 K in the Hg-based $n=3$ compound as a result of optimum doping at the OP and IPs.\cite{Schilling} In this context, it is remarkable that the dome shape of $T_c$ from the underdoped to the overdoped regime with a maximum $T_c$ at $N_h\sim$ 0.22 does not always depend on $n$, but its maximum value of $T_c$ seems to depend on $n$ moderately.
    
In light of the recent progress in these experiments unraveling the intrinsic phase diagrams in the underdoped regime, 
in this note, we deal with a carrier-density $N_h$ dependence of AFM moment and an in-plane superexchange $J_{\rm in}$ in order to gain insight into magnetic characteristics in the antiferromagnetic HTSC. 
\begin{figure}[h]
\centering
\includegraphics[width=9cm]{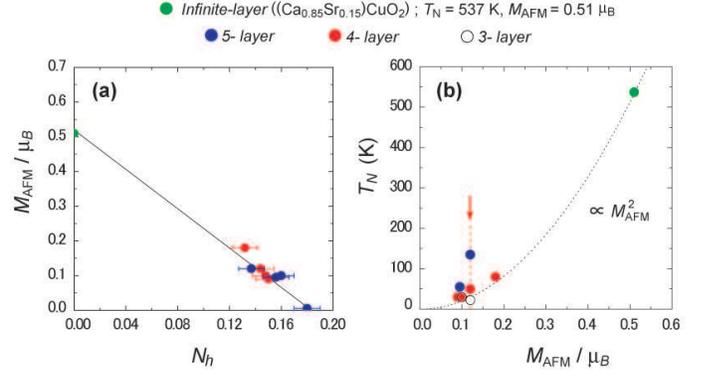}
\caption[]{\footnotesize (Color online) (a) Plot of $M_{\rm AFM}$ versus $N_h$ \cite{Shimizu2,Mukuda2008,Shimizu4,Vaknin}.  Solid line indicates a relation of $M_{\rm AFM}=-3N_h+0.51$. 
(b) Plot of $T_N$ versus $M_{\rm AFM}$ \cite{Shimizu1,Shimizu2,Mukuda2008,Shimizu4,Vaknin}. Dotted curve shows $T_N\propto M_{\rm AFM}^2(J_{\rm in}J_{\rm out})^{1/2}$ with $T_N=537$ K and $M_{\rm AFM}=0.51\mu_B$ in Ca$_{0.85}$Sr$_{0.15}$CuO$_{2}$ \cite{Vaknin} where $n=\infty$. A red arrow points to $M_{\rm AFM}\sim 0.12$ $\mu_B$.}
\end{figure}

Figure 2(a) shows a plot of AFM moment $M_{\rm AFM}$ versus $N_h$ where the datum at $N_h$=0 for a Mott insulator is cited from an infinite-layer compound Ca$_{0.85}$Sr$_{0.15}$CuO$_{2}$ (green circle) with $T_N=537$ K and $M_{\rm AFM}=0.51$ $\mu_B$.\cite{Vaknin} All other data are plotted with respect to those in the $n=4$ compounds (red circle) \cite{Shimizu2} and the $n=5$ compounds (blue circle).\cite{Mukuda2008,Shimizu4} Remarkably, $M_{\rm AFM}$ decreases linearly as the function of $N_h$ irrespective of $n$ with a relation of $M_{\rm AFM}=-3N_h+0.51$, when doped CuO$_2$ planes are magnetically ordered. A critical carrier density $N_h$(experiment) $\sim$ 0.17 is larger than a theoretical value $N_h$(theory) $\sim$ 0.10 for the $T=0$ phase diagram in a single CuO$_2$ plane where no long-range magnetic order takes place at a finite temperature.\cite{Giamarchi,Himeda,Paramekanti1,Paramekanti2,Chen,Lee1,Shih1,Shih2,Kotliar,Senechal,Pathak,Ogata,Kyung} We consider that a reason why $N_h$(theory) $\sim$ 0.10 is significantly smaller than $N_h$(experiment) $\sim$ 0.17 is because $J_{\rm out}$ responsible for the AFM order is not taken into account in the theories at all. 

In order to gain further insight into a $N_h$ dependence of $J_{\rm in}$, Fig.2(b) shows a plot of $T_N$ versus $M_{\rm AFM}$ where the data are presented with respect to Ca$_{0.85}$Sr$_{0.15}$CuO$_{2}$ for $N_h$=0(green circle),\cite{Vaknin} a $n=3$ compound (open circle),\cite{Shimizu1} the $n=4$ (red circle),\cite{Shimizu2} and the $n=5$ compounds (blue circle).\cite{Mukuda2008,Shimizu4} On the basis of the mean-field approximation of localized spins, $T_N$ is nearly proportional to $M_{\rm AFM}^2$ when assuming that $J_{\rm out}$ and $J_{in}$ stay constant regardless of $N_h$. The dotted curve in Fig.2(b) shows $T_N\propto M_{\rm AFM}^2(J_{\rm in}J_{\rm out})^{1/2}$ with $T_N=537$ K and $M_{\rm AFM}=0.51$ $\mu_B$ in Ca$_{0.85}$Sr$_{0.15}$CuO$_{2}$.  When noting that $J_{\rm out}$s for the $n=3$, 4, and 5 compounds become always smaller than the $J_{\rm out}$ in Ca$_{0.85}$Sr$_{0.15}$CuO$_{2}$ where $n=\infty$, an unexpected fact that most of the data are larger than those that would be expected from the dotted curve reveals that $J_{\rm in}\sim$~1300 K is not decreased by doping hole carriers, but it is even increased. Another important outcome extracted from Fig.2(b) is that even though $M_{\rm AFM}\sim 0.12$ $\mu_B$ is the same as shown by a red arrow, $T_N$ increases due to the increase of $J_{\rm out}$ as $n$ increases from $n=3$ to 5. The two experimental relationships, the plot of $M_{\rm AFM}$ versus $N_h$ shown in Fig.2(a) and the plot of $T_N$ versus $M_{\rm AFM}$ in Fig.2(b), suggest that the AFM ground state in the homogeneously doped CuO$_2$ layers is determined by $N_h$ and $J_{\rm out}$. It is surprising that the superexchange interaction $J_{\rm in}$ does not depend on $N_h$ so much, but is even increased. In HTSC, mean-field theories used to take the Heisenberg superexchange $J_{\rm in}$ as the source
of an instantaneous attraction that leads to pairing in a d-wave state.\cite{Anderson2} The present outcomes may support such a picture experimentally as far as the underdoped region with $N_h < 0.17$ is concerned where AFM and SC uniformly coexist in a CuO$_2$ plane.

In conclusion, on the basis of the extensive experimental works on the multilayered compounds,\cite{Mukuda2006,Mukuda2008,Shimizu,Shimizu1,Shimizu2,Shimizu3} we have presented the following outcomes;
\begin{enumerate}
\item $M_{\rm AFM}$ is decreased with doping, regardless of the number of layers $n$, and collapses around $N_h\sim$ 0.17.
\item The AFM ordering temperature is enhanced as the out-of-plane coupling $J_{\rm out}$ increases with increasing $n$.
\item The in-plane superexchange $J_{\rm in}$ is invariant with doping, but is even increased.
\item The dome shape of $T_c$ from the underdoped to the overdoped regime with a maximum $T_c$ at $N_h\sim$ 0.22 does not depend on $n$, but its maximum value of $T_c$ seems to depend on $n$ moderately.
\end{enumerate}
When noting that $T_c$ is maximum close to the QCP, the results presented here strongly suggest that the AFM interaction plays the vital role as the glue for the Cooper pairs, which will lead us to a genuine understanding of why the $T_c$ of cuprate superconductors is so high. In fact, we note that a recent theoretical analysis based on a cellular dynamical mean-field theory of Hubbard model has revealed that an energy scale in spin-fluctuations spectrum that leads to pair binding is of order of the Heisenberg superexchange $J_{\rm in}$ independent of doping.\cite{Kyung}

We are grateful to P.M. Shirage, H. Kito, Y. Kodama and A. Iyo for providing the samples. This work was supported by a Grant-in-Aid for Specially Promoted Research (20001004) and by Global COE Program (MEXT).

\end{document}